\documentclass[aps,prapplied,twocolumn,superscriptaddress]{revtex4-2}

\usepackage{graphicx}
\usepackage{dcolumn}
\usepackage{bm}
\usepackage{booktabs}
\usepackage{siunitx}
\usepackage[labelformat=simple,caption=false]{subfig}

\usepackage{xcolor}

\newcommand{\beginsupplement}{
  \newcounter{offset}
  \setcounter{offset}{\value{figure}}
  \renewcommand{\thefigure}{S\the\numexpr\value{figure}-\value{offset}\relax}
}

\begin{document}

\title{Fast optical data transfer into a Josephson junction array}

\author{K. Kohopää}
\email{katja.kohopaa@vtt.fi}
\thanks{\\
\copyright\ 2026 American Physical Society. This is the accepted manuscript of the following article: K. Kohopää et al., "Fast optical data transfer into a Josephson-junction array", Phys. Rev. Appl. (2026). The final published version is available from https://doi.org/10.1103/tql5-gt9g. This manuscript version is posted with permission for non-commercial scholarly use.}
\affiliation{VTT Technical Research Centre of Finland Ltd, P.O. Box 1000, FI-02044 VTT, Finland}
\author{J. Nissilä}
\affiliation{VTT Technical Research Centre of Finland Ltd, P.O. Box 1000, FI-02044 VTT, Finland}
\author{E. Mykkänen}
\affiliation{VTT Technical Research Centre of Finland Ltd, P.O. Box 1000, FI-02044 VTT, Finland}
\author{P. Selvasundaram}
\affiliation{VTT Technical Research Centre of Finland Ltd, P.O. Box 1000, FI-02044 VTT, Finland}
\author{T. Fordell}
\affiliation{VTT Technical Research Centre of Finland Ltd, P.O. Box 1000, FI-02044 VTT, Finland}
\author{K. Langi}
\affiliation{VTT Technical Research Centre of Finland Ltd, P.O. Box 1000, FI-02044 VTT, Finland}
\author{E. T. Mannila}
\affiliation{VTT Technical Research Centre of Finland Ltd, P.O. Box 1000, FI-02044 VTT, Finland}
\author{S. Kafanov}
\affiliation{Department of Physics, Lancaster University, Lancaster LA1 4YB, United Kingdom
}
\author{S. Ahopelto}
\affiliation{VTT Technical Research Centre of Finland Ltd, P.O. Box 1000, FI-02044 VTT, Finland}
\author{H. Systä}
\affiliation{VTT Technical Research Centre of Finland Ltd, P.O. Box 1000, FI-02044 VTT, Finland}
\author{M. Ribeiro}
\affiliation{VTT Technical Research Centre of Finland Ltd, P.O. Box 1000, FI-02044 VTT, Finland}
\affiliation{Present address: Arctic Instruments Oy, Tekniikantie 14, 02150 Espoo, Finland}
\author{P. Sethi}
\affiliation{VTT Technical Research Centre of Finland Ltd, P.O. Box 1000, FI-02044 VTT, Finland}
\author{M. Kiviranta}
\affiliation{VTT Technical Research Centre of Finland Ltd, P.O. Box 1000, FI-02044 VTT, Finland}
\author{R. Loreto}
\affiliation{VTT Technical Research Centre of Finland Ltd, P.O. Box 1000, FI-02044 VTT, Finland}
\author{J.-W. Lee}
\affiliation{VTT Technical Research Centre of Finland Ltd, P.O. Box 1000, FI-02044 VTT, Finland}
\author{T. Rantanen}
\affiliation{VTT Technical Research Centre of Finland Ltd, P.O. Box 1000, FI-02044 VTT, Finland}
\author{V. Vesterinen}
\affiliation{VTT Technical Research Centre of Finland Ltd, P.O. Box 1000, FI-02044 VTT, Finland}
\affiliation{Present address: Arctic Instruments Oy, Tekniikantie 14, 02150 Espoo, Finland}
\author{O. Kieler}
\affiliation{Physikalisch-Technische Bundesanstalt, Bundesallee 100, 38116 Braunschweig, Germany}
\author{M. Bieler}
\affiliation{Physikalisch-Technische Bundesanstalt, Bundesallee 100, 38116 Braunschweig, Germany}
\author{J. Govenius}
\affiliation{VTT Technical Research Centre of Finland Ltd, P.O. Box 1000, FI-02044 VTT, Finland}
\affiliation{Present address: Arctic Instruments Oy, Tekniikantie 14, 02150 Espoo, Finland}
\author{J. Senior}
\affiliation{VTT Technical Research Centre of Finland Ltd, P.O. Box 1000, FI-02044 VTT, Finland}
\author{A. Kemppinen}
\affiliation{VTT Technical Research Centre of Finland Ltd, P.O. Box 1000, FI-02044 VTT, Finland}

\date{\today}

\begin{abstract}
We employ externally shunted Nb--AlO$_x$--Nb Josephson junctions for demonstrating a circuit that is suitable for an optically driven Josephson Arbitrary Waveform Synthesizer (JAWS). This technology enables overdamped junctions with characteristic frequencies above 100 GHz and critical currents of the order of 100~µA, which is promising, e.g., for low-dissipation optical control of quantum circuits such as superconducting quantum bits. Here we utilize a double-pulse technique to experimentally determine the maximum rate at which optical pulse data can be reliably delivered to the superconducting circuit. We demonstrate the feasibility of data transfer up to 60~Gbit/s, which is about factor 4 higher than for typical JAWS.
\end{abstract}

\maketitle

Superconducting quantum computers are leading the way in scaling up cryogenic technologies, approaching 1000 quantum bits (qubits), and thus stretching the limits of refrigeration technology~\cite{raicu_cryogenic_2025}. However, fault-tolerant, universal quantum computers may require up to a million or more physical qubits, which poses a major challenge for their energy-efficiency, both due to their technical feasibility~\cite{fowler_surface_2012, mohseni_how_2025} and for enabling sustainable quantum computing~\cite{arora_sustainable_2024}. The energy consumption of a cryogenic assembly is typically dominated by that of its refrigerator system, which scales by the heat load at low temperatures. The load consists of two main components: passive load arising from heat conduction from room temperature into the cryogenic system and active load due to dissipation. The significant passive load, mainly due to the thermal conductivity of microwave coaxial cables, is in principle straightforward to suppress by replacing cables with optical fibers and driving qubits with photodiodes. However, this direct optical--to--microwave transduction leads to a tradeoff between shot noise and energy efficiency~\cite{lecocq_control_2021}. It is thus crucial to develop methods for optics-based cryogenic signal generation that, in addition to suppressing the passive heat load, also enable minimal dissipation at cryogenic temperatures. Superconducting circuits like Single Flux Quantum (SFQ) technology~\cite{likharev_rsfq_1991} and many of its variants are capable of generating qubit drive signals based on digital data~\cite{mcdermott_accurate_2014, liebermann_optimal_2016, mcdermott_quantumclassical_2018, leonard_digital_2019, li_hardware-efficient_2019,liu_single_2023}. Since digital processing reduces noise sensitivity, using optical digital data transfer instead of analogue signal transduction may alleviate the tradeoff between signal quality and energy consumption.

In this letter, we focus on fast optical data transfer into Josephson Arbitrary Waveform Synthesizers (JAWS)~\cite{benz_pulsedriven_1996, kieler_sns_2007, urano_pulse_2008, nissila_driving_2021}. Their optical control requires first a digital pulse pattern consisting of existing or missing optical pulses, which are transduced into electrical current pulses with a photodetector. A Josephson junction array (JJA) quantizes each current pulse such that the average voltage across the JJA equals $V=Ns(t) \Phi_0 f_\mathrm{pulse}(t)$. Here, $N$ is the number of Josephson junctions (JJ) in the JJA, $s(t)$ is an integer describing the Shapiro step index to which the array is driven at time $t$, $f_\mathrm{pulse}(t)$ is the time-dependent pulse frequency, and $\Phi_0=h/(2e)$ is the magnetic flux quantum, where $h$ is the Planck constant and $e$ is the elementary charge.

In comparison to the classical computing technology SFQ, JAWS is instead a digital-to-analogue converter capable of generating quantized voltage signals. Furthermore, JAWS enables a wider range of voltages (in SFQ, $N=1$) and especially when driven optically, potentially excellent timing jitter~\cite{nissila_driving_2021}. While both of these JJ-based circuits are in principle equally compatible with high pulse frequencies $f_\mathrm{pulse}>100$~GHz (in other words: clock frequency, data rate), they have only been demonstrated for SFQ~\cite{likharev_rsfq_1991}. Traditional electrical control technology has limited JAWS to about $f_\mathrm{pulse}=15$~GHz~\cite{kieler_optical_2019,babenko_microwave_2021,benz_ac_2024} whereas the optical control is in the early stages of development~\cite{nissila_driving_2021, brevik_bipolar_2022, lee_sub-ghz_2023}. The control of qubits has only been demonstrated for a simplified variant of JAWS, called Josephson Pulse Generator~\cite{howe_digital_2022, castellanos-beltran_coherence-limited_2023}, whereas generating qubit drive signals with JAWS is difficult due to the slow data rate~\cite{sirois_josephson_2020, donnelly_thesis_2019}. However, compared to SFQ, JAWS can produce substantial signals that can be attenuated and filtered for qubit control, enabling operation at elevated temperatures and avoiding problems with quasiparticle poisoning~\cite{howe_digital_2022}. In this letter, we study optical control as a method to improve the energy efficiency and pulse frequency of JAWS, which we consider as critical steps toward scalable qubit control. Actual experiments with qubits still require significant further engineering effort including design and realization of synchronizable gate electronics at room temperature for the optical pulse pattern generation~\cite{nissila_driving_2021}, but we do not see any fundamental issues on that path.

The correct operation of the JAWS requires overdamped JJs, for which the Stewart--McCumber parameter $\beta_c = (f_c/f_p)^2 < 1$. Here, $f_c=I_cR/\Phi_0$ and
$f_p=\sqrt{I_c/(2\pi \Phi_0 C)}$ are the characteristic and plasma frequencies of the JJs, respectively, and $I_c$, $R$, and $C$ are the critical current, shunt resistance and capacitance of the JJs, respectively. The characteristic frequency also determines the maximum rate of pulse data, i.e., $f_\mathrm{pulse}\lesssim f_c$. Conventional JAWS technology typically utilizes self-shunted junctions, e.g., superconductor -- normal metal -- superconductor (SNS) technology~\cite{benz_pulsedriven_1996,kieler_sns_2007}, which enables high-density JJAs and small, m$\Omega$ range $R$. Since the total resistance of the JJA, $R_\mathrm{JJA}=NR$, should not exceed the transmission line impedance, the SNS technology enables large number of junctions $N$ and critical current $I_c$. For the traditional use of JAWS, i.e., metrological calibrations of room temperature electronics, these are beneficial properties since large $N>1000$ increases the maximum output voltage whereas substantial $I_c$ yields robustness against noise of room-temperature electronics. However, scaling up quantum technologies through optical control sets significantly different criteria: a small $I_c$ minimizes dissipation~\cite{nissila_driving_2021}, whereas a large characteristic frequency $f_c\propto I_cR$ increases the bandwidth. This favors increasing $R$, which yields a smaller maximum value of $N$, but on the other hand, the required output signal levels are usually small. Another characteristic of SNS junctions is that their properties typically have a strong temperature dependence~\cite{baek_co-sputtered_2006,haygood_characterization_2019}.

\begin{figure}[!t]
    \includegraphics[width=\columnwidth]{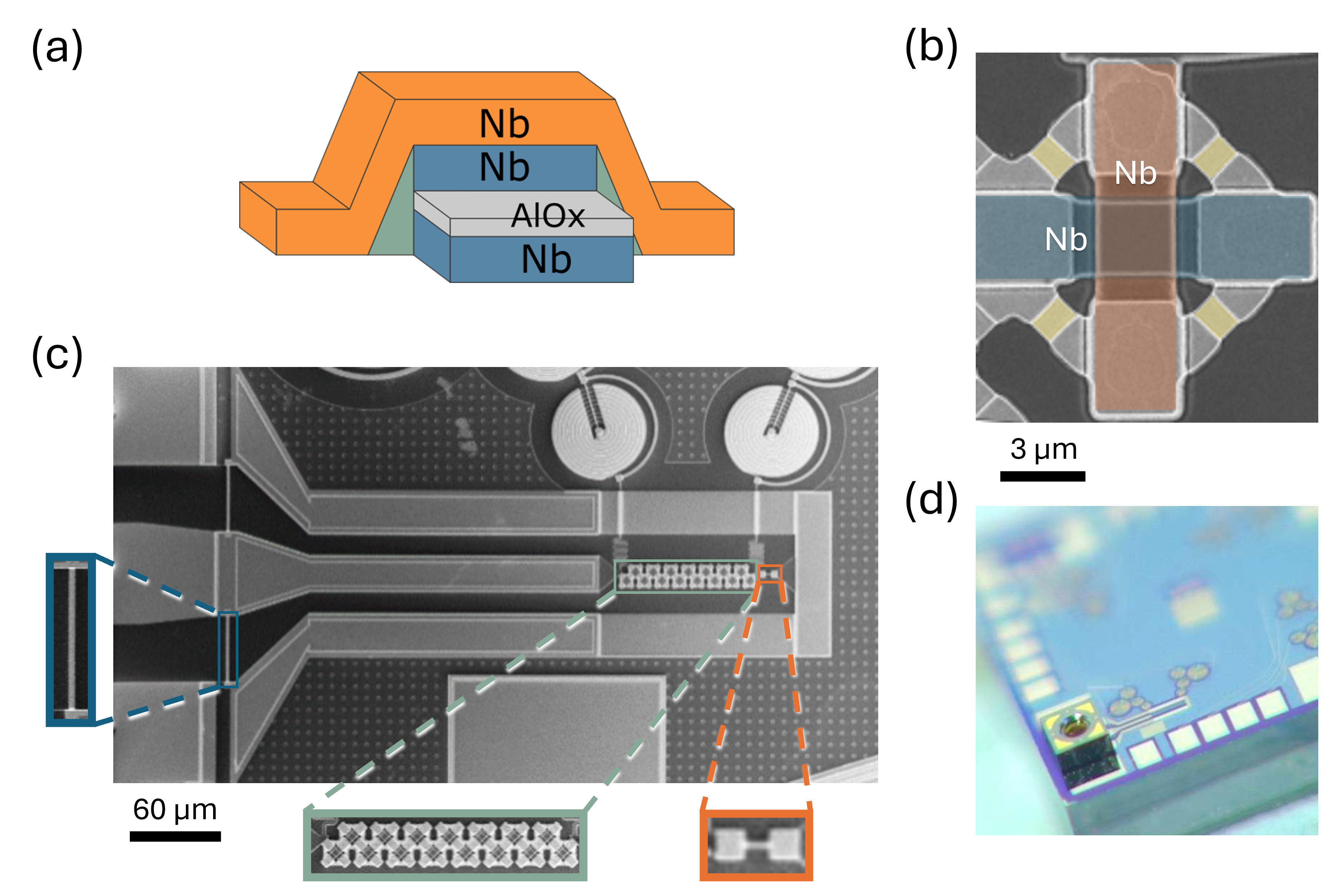}
    \caption{(a) Schematic of cross-cut layer structure of niobium SIS junctions. The green triangular structure is silicon oxide sidewall passivation that prevents contact between the orange wiring layer of the top electrode of the JJ and the bottom electrode. Thus, the JJ area is determined only by the blue--gray Nb--AlO$_x$--Nb trilayer structure. (b) SEM image of a single Josephson junction, in which the two Nb layers are highlighted with blue and orange. To minimize parasitic inductance, the shunt resistor consists of four parallel resistors of $4R$ surrounding the junction. The resistors are fabricated from TiW and are highlighted in yellow. (c) SEM image of a similar JAWS chip as the measured one. On the left, there are 3 flip-chip bonding pads separated by a gap that yields a \SI{50}{\ohm} CPW, which is tapered down to narrower dimensions. The CPW delivers photodiode signals to the JJA highlighted in green. The input and termination resistors are also highlighted (in blue and orange, respectively). The large inductors at the top part of the image are parts of low-pass filters, see Fig.~\ref{fig:fig2}(b). (d) Photodiode flip-chip bonded on a JAWS chip.}
    \label{fig:fig1}
\end{figure}

Here we demonstrate JAWS chips based on externally shunted superconductor--insulator--superconductor (SIS) junctions. They allow low $I_c$ and large $f_c$, which may in the future enable energy-efficient control of qubits, see the Supplemental Material~\cite{supplementary} for discussion. Figure~\ref{fig:fig1}(a) shows the cross-cut structure of our SIS junctions fabricated using sidewall-passivating spacer structure (SWAPS) process~\cite{gronberg_side-wall_2017}, based on niobium trilayers. Figures~\ref{fig:fig1}(b) and \ref{fig:fig1}(c) show scanning electron microscope (SEM) images of a single JJ and the most essential elements of the superconducting circuit, respectively. The circuit consists of flip-chip bonding pads for mounting the photodiode, a coplanar waveguide (CPW) designed for \SI{50}{\ohm} impedance that transmits signals to the JJA, and filters allowing both dc current bias and signal output. To mitigate the risks of novel technology, we did not yet optimize these filters for signal output in the GHz range, which is typical for driving superconducting qubits. Here we focus on a sample with a JJA consisting of 15 junctions with $I_c\approx$~\SI{170}{\uA} and $R\approx$~\SI{1.5}{\ohm}, giving to $f_c\approx$ \SI{125}{GHz}. The array was terminated with a resistor $R_\mathrm{end}\approx$~\SI{23}{\ohm} (highlighted in orange). The total transmission line termination resistance was $R_\mathrm{JJA}+R_\mathrm{end}\approx$~\SI{46}{\ohm}. To mitigate potential transmission line mismatches, e.g., due to fabrication scatter, our circuit also includes two symmetrically placed shunt resistors near the photodiode (highlighted in blue in Fig.~\ref{fig:fig1}(c)), yielding $R_\mathrm{in}\approx$~\SI{51}{\ohm}. See the Supplemental Material~\cite{supplementary} for details of the experimental determination of these parameters.

To drive the JJA optically, we use a commercial, nominally $\SI{60}{GHz}$ InGaAs photodiode (PD) from Albis Optoelectronics~\cite{Albis_PD}. Recent electro-optic sampling experiments~\cite{priyadarshi_-situ_2025} indicate that when driven with picosecond-range optical pulses in cryogenic temperatures, the bandwidth of these PDs may exceed \SI{120}{GHz}, i.e., it can be in the same range as $f_c$ of our JJA. Figure \ref{fig:fig1}(d) shows an Albis photodiode flip-chip bonded on a JAWS chip. On top of these two chips, we placed a perforated glass chip to allow mounting a polarization-maintaining single-mode fiber on top of the photodiode. The fiber was glued to the glass chip. The whole assembly was placed in a liquid helium dipstick, which thermalizes the sample assembly to \SI{4.2}{\K}.

\begin{figure}[!t]
    \includegraphics[width=\columnwidth]{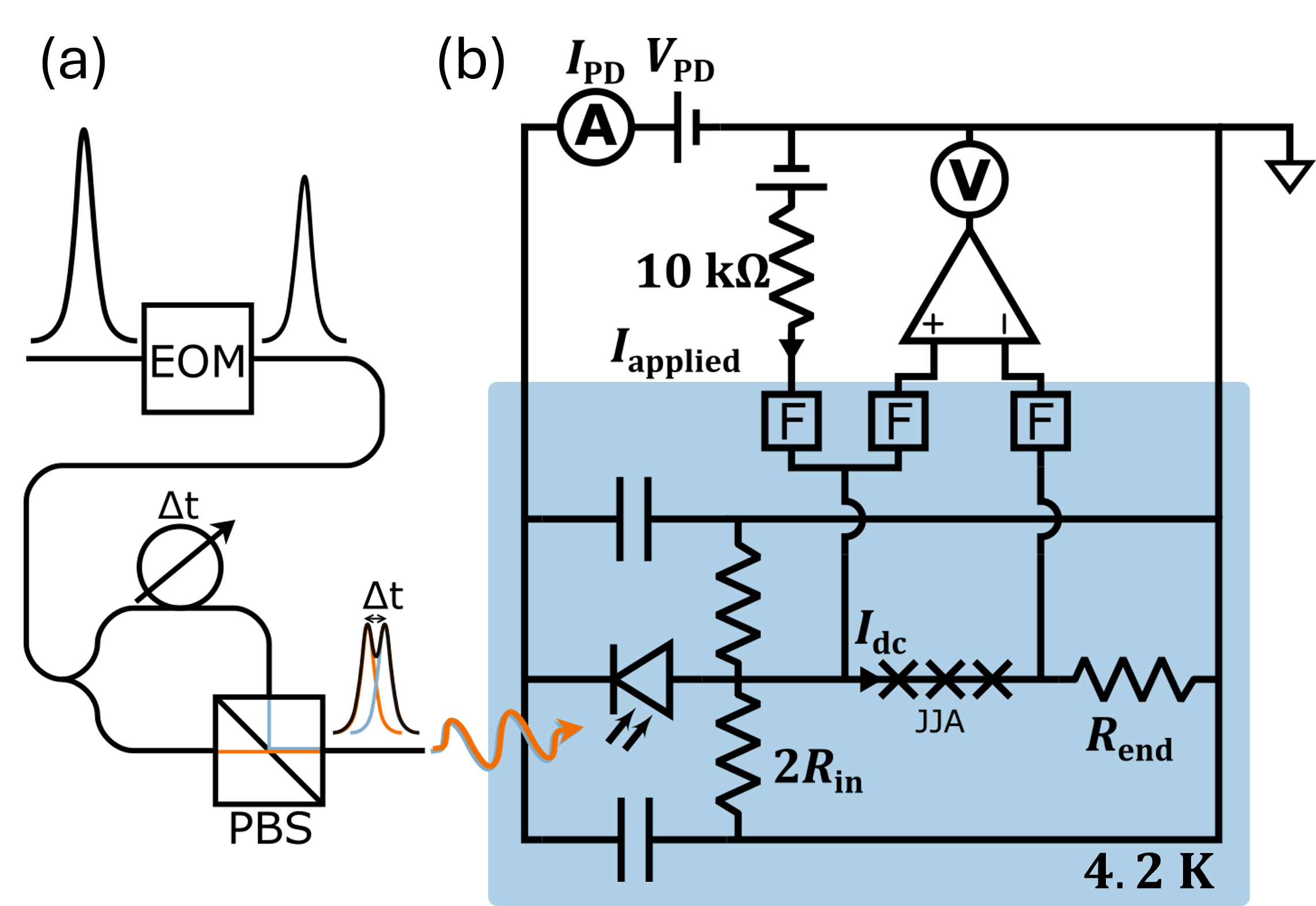}
    \caption{(a) Depiction of producing double pulses. Here, EOM varies the amplitude of pulses, optical fiber splitter divides them into two paths with variable travel time difference $\Delta t$, and PBS collects the pulses back to a polarization-maintaining fiber coupled to the photodiode. (b) Circuit diagram of the measurement setup. Blue area represents the sample that was at \SI{4}{\kelvin}, whereas the other parts on white background represent the measurement equipment at room temperature. In room temperature, we have photodiode voltage bias $V_\mathrm{PD}$, photocurrent measurement $I_\mathrm{PD}$ and dc current source $I_\mathrm{applied}$ which consists of voltage source and \SI{10}{k\ohm} resistor. A differential voltage amplifier is used in measuring the voltage over the JJA. In the sample, $R_\mathrm{in} = \SI{51}{\ohm}$ and $R_\mathrm{end} = \SI{23}{\ohm}$ are the input and termination resistors, respectively. The low-pass filters are presented by F. }
    \label{fig:fig2}
\end{figure}

Circuit diagram of the sample and the measurement setup is shown in Fig.~\ref{fig:fig2}(b). We measure the average voltage over the JJA and the average photocurrent $I_\mathrm{PD}$ of the diode while varying the optical pulse signal to the PD and the dc current $I_\mathrm{dc}$ through the JJA. To minimize the risk of damaging the PD, we kept its reverse voltage bias at \SI{1}{\V}, even though higher values may improve its bandwidth and linearity~\cite{priyadarshi_-situ_2025}. To allow for biasing the PD, both ground arms of the CPW were galvanically isolated by capacitors close to the PD. These capacitors stored the charge of about 200~pC, which is much larger than that of any current pulses in our experiments. Thus, we expect no bias droop due to inductance of the PD biasing wires.

Previous experiments on optical data transfer into JAWS \cite{nissila_driving_2021} have shown that transmission line reflections between the PD and the JJA may cause significant limitations for the reliable data transfer into the JJA. However, those experiments were performed with 20 GHz PDs and JJAs with $f_c<10$~GHz that were located centimeters away from each other and connected together by multiple interfaces. In our present experiments, we expect such problems to be significantly mitigated by flip-chip bonding the PD directly onto the JAWS chip and thus realizing a nearly ideal transmission line from PD to JJA. However, the present PD and JJA have a significantly higher bandwidth, and JJA as a nonlinear component is inherently difficult to impedance match. Thus, it is important to evaluate the data transfer bandwidth experimentally and quantitatively.

To enable this, we utilize the double pulse method \cite{nissila_driving_2021}: We drive the JAWS with two orthogonally polarized optical pulses, illustrated in Fig.~\ref{fig:fig2}(a). We initially generate pulses at \SI{1310}{\nm} wavelength and repetition rate of $f_\mathrm{period}=\SI{250}{\MHz}$ with a commercial frequency comb. By utilizing chromatic dispersion of optical fibers, we obtain pulses with time duration of about 2 ps (full width at half maximum, FWHM), suitable for driving the PDs~\cite{priyadarshi_-situ_2025}. The initial pulses are linearly polarized and an electro-optic modulator (EOM) is used to tune their amplitude. Each pulse is split into two pulses of equal energy, but with orthogonal polarizations. We control the time difference $\Delta t$ between the two pulses by varying the free space travel distance of one of them with a translation stage. Details on determining the resulting values of $\Delta t$ are provided in the Supplemental Material~\cite{supplementary}. The pulses are combined in a single optical fiber by utilizing a polarizing beam splitter (PBS). The orthogonal polarization of the two pulses in the pair allows temporal and spatial overlap without interference.

During measurements, we vary $\Delta t$, the magnitude of the current pulses, and the applied dc bias current $I_\mathrm{applied}$. Due to the input resistors of our circuit (see Fig.~\ref{fig:fig2}(b)), the currents have two routes. We define $I_\mathrm{dc}$ of the JJA assuming that the JJs are in the superconducting state, i.e., utilizing the current division between $R_\mathrm{in}$ and $R_\mathrm{end}$ that yields $I_\mathrm{dc}\approx 0.7\times I_\mathrm{applied}$. In contrast, we can only measure the total current produced by the PD, giving the total time integral of pulse currents over one period, $Q_\mathrm{PD}=I_\mathrm{PD}/f_\mathrm{period}$. Its division between $R_\mathrm{in}$ and JJA is non-trivial due to the nonlinear impedance of the JJs, see the Supplemental Material~\cite{supplementary} for more details. 

We measure the average voltage $V$ over multiple periods of optical pulses. We normalize the resulting voltage as $v = V / (N \Phi_0 f_\mathrm{period}) \simeq V / (7.8~\mathrm{\mu V})$. Since there are two pulses per period, for large $\Delta t$ both of them yield the same Shapiro step, and only even plateaus with $v=0,2,4\ldots$ can be observed. In the other extreme, at $\Delta t\approx 0$, the two pulses overlap and are indistinguishable by the JJA. Then the two pulses simply sum up, yielding all integer value plateaus of $v$. Studying the crossover between these two extremes allows a quantitative measurement of the bandwidth of the system, i.e., the minimum $\Delta t$ at which the JAWS only yields the even steps of $v$.

\begin{figure}[!t]
    \includegraphics[width=\columnwidth]{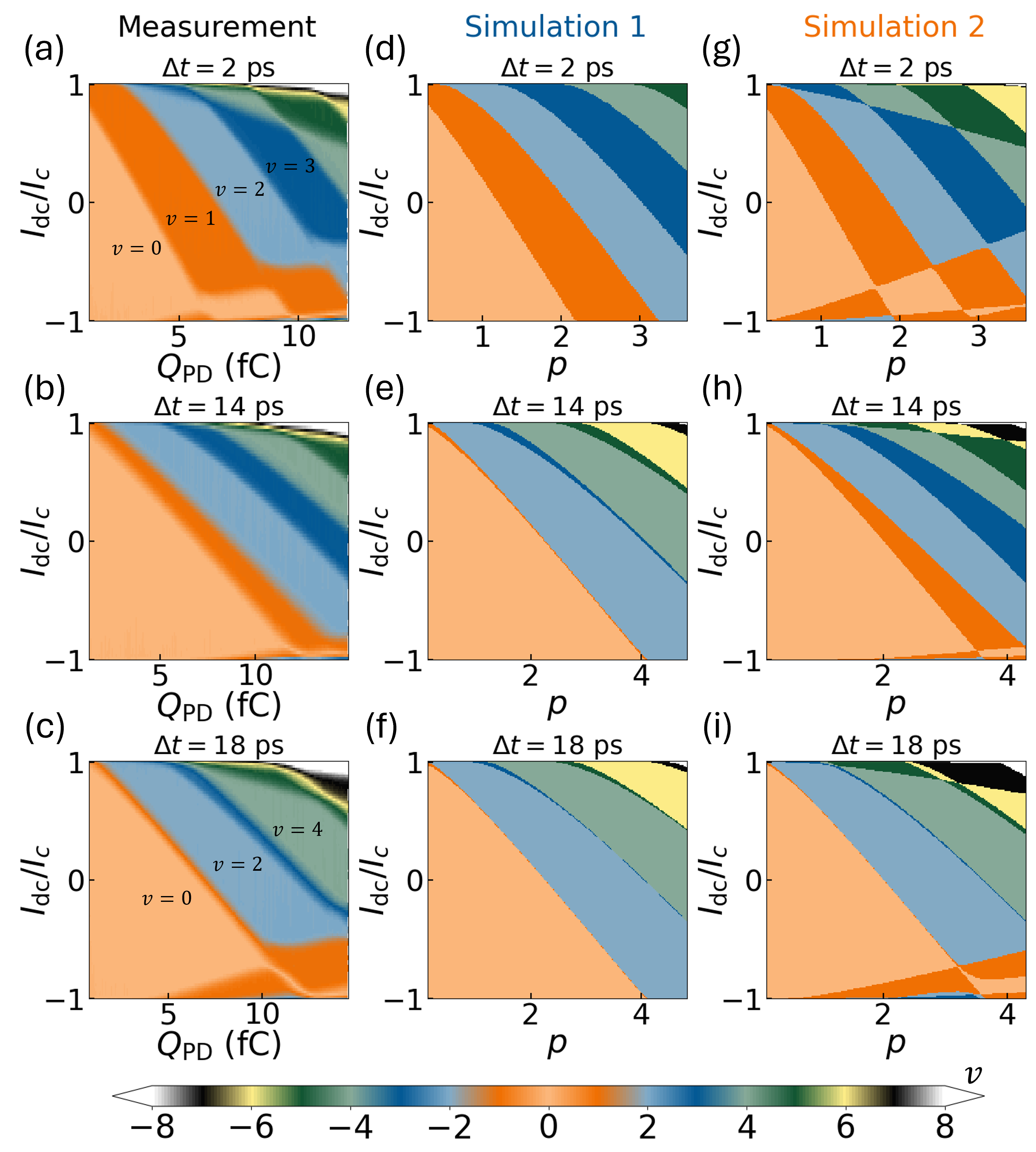}
    \caption{Measurements and simulations of double pulse experiments with $\Delta t = $ 2~ps, 14~ps and 18~ps. (a--c) Measurements of $v$ as the function of $I_\mathrm{dc}/I_c$ and $Q_\mathrm{PD}$. (d--f) and (g--i) Simulation model 1 and 2 results, respectively, corresponding to the experiments of panels (a--c). The X axes of the simulation results show the normalized pulse integrals, which are expected to roughly correspond to $Q_\mathrm{PD}$ with an unknown scaling factor, see the Supplemental Material~\cite{supplementary} for details.}
    \label{fig:fig3}

\end{figure}

Figures~\ref{fig:fig3}(a--c) show results of the double pulse experiments as the function of $Q_\mathrm{PD}$ and normalized dc current $I_\mathrm{dc}/I_c$ for three values of $\Delta t$. In Fig.~\ref{fig:fig3}(a), each voltage plateau has roughly equal width along the $Q_\mathrm{PD}$ axis at $I_\mathrm{dc}/I_c=0$. This is expected since $\Delta t=2$~ps is clearly below the expected characteristic timescales of both PD and JJA ($1/f_c\approx 8$~ps) and thus the two pulses are indistinguishable. In Fig.~\ref{fig:fig3}(c), at $\Delta t = 18$~ps, the odd plateaus are negligible, which is a sign of clearly distinguishable pulses and demonstrates reliable data transfer for this pulse separation. However, Fig.~\ref{fig:fig3}(b) ($\Delta t=14$~ps) is in the crossover regime where the odd plateaus are significant, but narrower than the even ones. This is an indication that the response to bit 1 (existing pulse) of a JAWS sequence is not completely independent of whether it was preceded by bit 0 or by bit 1. Thus, the margins for reliable data transfer are narrower than ideally. This result cannot be explained by the bandwidths of the PD and JJA only.

Figures~\ref{fig:fig3}(d--f) show results of our simplest simulation model, called simulation 1, corresponding to the experimental data of Figs.~\ref{fig:fig3}(a--c). Here, we apply Gaussian pulses of width 5~ps (FWHM) from an ideal current source into a single shunted JJ, neglecting the other resistors, the output filters and the CPW. The division of $Q_\mathrm{PD}$ cannot be described with this model, and thus we report the simulation data instead as the function of the normalized current pulse integral $p = \frac{f_c}{I_c} \int_0^{1/f_\mathrm{period}} I_p(t) \mathrm{dt} = f_c I_\mathrm{PD}/(I_c f_\mathrm{period})$. Here, $I_p(t)$ is the pulse current through the JJ. Already this simple model yields a relatively good qualitative description of the measurement results but with two significant differences: the model predicts vanishing odd plateaus already for $\Delta t = 14$~ps (Figs.~\ref{fig:fig3}(b) and \ref{fig:fig3}(e)) and it does not describe features of the experimental data with $|I_\mathrm{dc}|$ approaching $I_c$ especially for the higher $p$. We note that we do not know precisely the width of the pulses at 4~K during the measurements, and the mismatch between experimental and simulated plateau widths at $\Delta t = 14$~ps can be mitigated by simulating with unexpectedly long pulses as shown in the Supplemental Material~\cite{supplementary}. However, such simulations do not explain the other discrepant features.

\begin{figure}[!t]
    \includegraphics[width=\columnwidth]{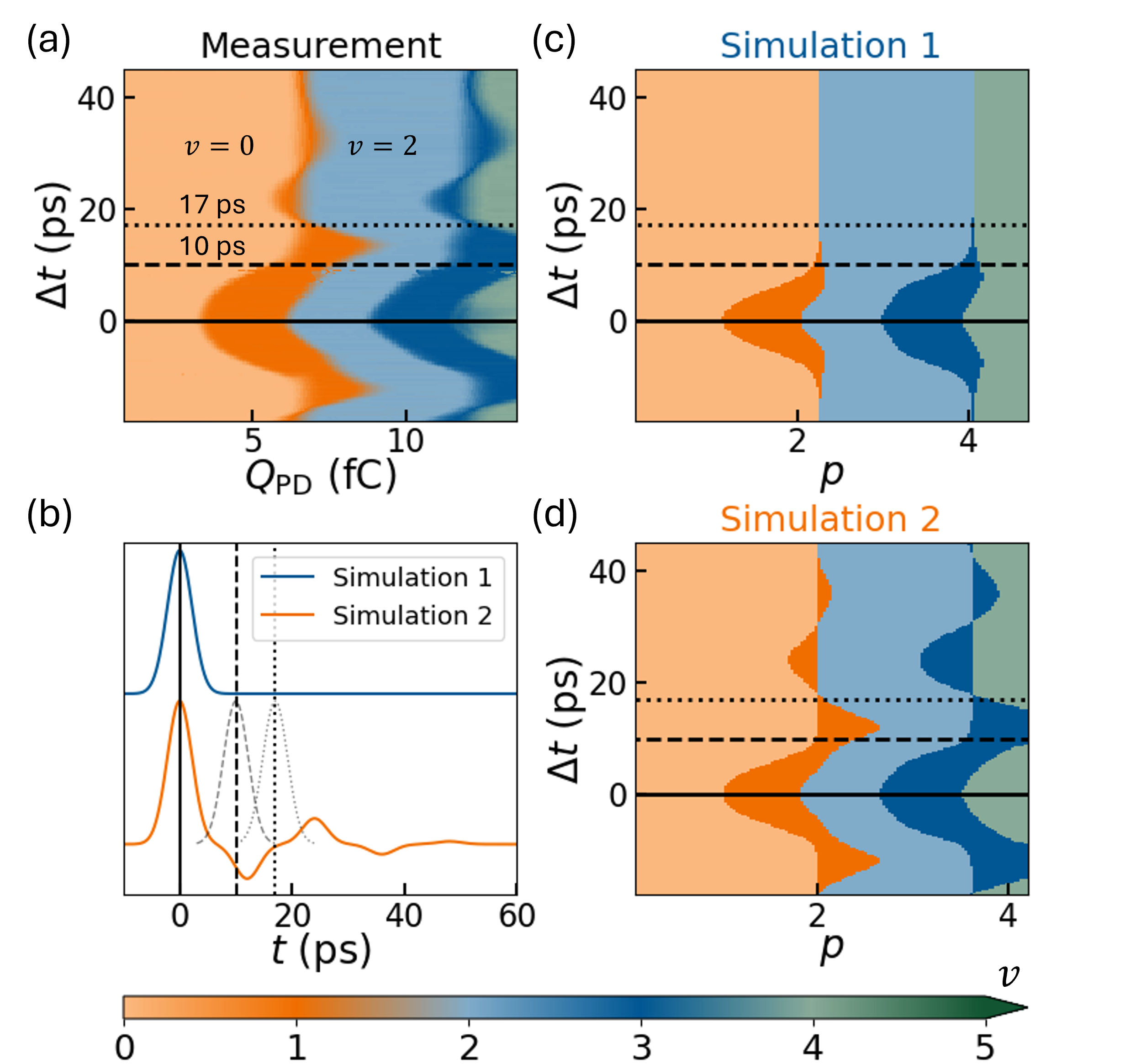}
    \caption{(a) Normalized voltage $v$ as a function of $\Delta t$ and $Q_\mathrm{PD}$ for $I_\mathrm{dc}=0$. (b) Individual current pulse shapes of simulations 1 and 2 (blue and orange, respectively). Grey dashed and dotted pulse shapes illustrate the position of the second pulse of the pair with $\Delta t = $ 10~ps and 17~ps, respectively. (c--d) Simulation model 1 and 2 results, respectively, corresponding to the experiments of panel (a).}
    \label{fig:fig4}

\end{figure}

Figure~\ref{fig:fig4}(a) shows the voltage plateaus at $I_\mathrm{dc} = 0$ when $\Delta t$ and $Q_\mathrm{PD}$ are varied. Figure~\ref{fig:fig4}(c) shows the corresponding result from simulation 1. While the simulation predicts negligible odd plateaus already at $\Delta t=\SI{10}{\ps}$, the experimental odd plateaus show dampening oscillatory features (around $Q_\mathrm{PD}\approx \SI{6}{\femto\coulomb}$ for $v=1$) up to $\Delta t\approx 40$~ps. However, beyond $\Delta t = 17$~ps, the width of odd plateaus becomes nonessential. These results thus imply that reliable data transfer into JAWS system is feasible up to about $1/\SI{17}{\ps} = \SI{60}{\GHz}$.

Previous experiments~\cite{nissila_driving_2021} and the oscillatory features of Fig.~\ref{fig:fig4}(a) suggest the possible presence of dampening transmission line reflections in our experiments. Potential origins for such reflections include the imperfect matching of transmission line termination resistances, the nonlinear properties of the JJA that affect the termination even with perfect resistance matching, and the propagation of the PD pulses into the filters, see Figs.~\ref{fig:fig1}(c) and \ref{fig:fig2}(b). Even though the filters were aimed to yield a high impedance for the pulse signals, thus allowing for reliable pulse propagation into the JJA, their design is non-trivial due to the inevitable parasitic capacitances and possible resonances arising from the resulting LC circuit. In addition to transmission line effects, it is also possible that the inductance of the PD (the exact design of which is unknown to us) may lead to a resonance with the capacitors of the ground arms.

To gain understanding to the oscillatory features, we first performed a phenomenological simulation (no.~2) that is otherwise similar as simulation 1, but instead of ideal current pulses, we assume small oscillations of the ideal current source, see Fig.~\ref{fig:fig4}(b). The time interval of the oscillations, 12~ps, and the amplitudes were estimated by matching the resulting simulation data to those of Fig.~\ref{fig:fig4}(a). This simple model yields an excellent fit to experimental data for both Figs.~\ref{fig:fig3} and~\ref{fig:fig4}. This simulation also suggests an explanation, why the $v=1$ plateau extends to high $Q_\mathrm{PD}$ between $\Delta t=10$ and 17~ps. In this range, the first negative phase of oscillation of the first pulse coincides with the second pulse, which therefore needs higher magnitude to yield the single $\Phi_0$ voltage pulse in the JJA. The 12~ps interval also roughly corresponds to the expected signal propagation time for traveling back and forth the about \SI{600}{\um} distance between the PD and the end of the JJA, whereas reflections between the two ends of the JJA would be much faster.

In the Supplemental Material~\cite{supplementary}, we provide more thorough simulations of the whole system. We performed simulations that included the transmission line, JJA and resistors, and those results are in general in good agreement with the simple simulations presented here, i.e., they did not reveal any significant transmission line reflections. We also did simulations with a simplified model of the filter coils which imply that they are a realistic source for the observed oscillations. A complete finite element method simulation of the filters and the rest of the JJ circuit is beyond the scope of the present work.

We envision that in the future, optical JAWS can be operated at, e.g., the 1~K or 4~K stage of a dilution refrigerator to drive qubits via attenuated microwave cables, see the Supplemental Material~\cite{supplementary} for, e.g., the required voltage levels and dissipation estimates. In the present experiment, $f_\mathrm{period}=250$~MHz yields negligible dissipation that therefore could not be measured directly. We estimate that with $f_\mathrm{pulse}=60$~GHz and an even distribution of bits 0 and 1, the present device would dissipate about 0.5~mW.

To conclude, we have demonstrated that JAWS can be driven optically with the pulse frequency of 60~GHz, which is about four times higher than in typical electrically driven JAWS. The externally-shunted SIS junctions presented here have high characteristic frequency of \SI{125}{GHz} and low $I_c$ of \SI{170}{\micro\ampere},  which is expected to enable the energy efficient generation of even higher frequencies in the future. Such improvements will, however, require improved transmission of ultrafast electrical signals to the Josephson junction array. For example, using cryogenic electro-optic sampling techniques~\cite{priyadarshi_cryogenic_2024,priyadarshi_-situ_2025} might shed light into the signal propagation problem. Another approach would be to eliminate the transmission line between the array and the photodiode, thus enabling an essentially lumped element system. Nevertheless, our results pave the way for ultrafast JAWS and using it to drive superconducting qubits, which may enable improving the energy efficiency of quantum computing.

\section*{Acknowledgements}

The authors thank Isabel Gueissaz-Mattelmäki for invaluable technical support. We thank Kirsi Tappura, Matteo Cherchi, Antti Manninen, Stefan Koepfli, Jukka Viheriälä, Gheorghe-Sorin Paraoanu, Marko Kuzmanovic and Ilya Moskalenko for fruitful discussions. 

We acknowledge the Research Council of Finland for funding through Grant Nos.~350220/QuantLearn, 359396/q.JAWS and 359284/Finnish Quantum Flagship. This work is part of the Research Council of Finland Flagship Programme "Photonics Research and Innovation" (PREIN, decision 320168). This work was also supported by the EMPIR programme co-financed by the Participating States and by the European Union’s Horizon 2020 Research and Innovation Programme under Grant Agreements 20FUN07/SuperQuant and 23FUN08/MetSuperQ, by the European Union’s Horizon 2020 and Horizon Europe programmes under Grant Agreements 899558/aCryComm, 101113946/OpenSuperQPlus100, 101113901/Qu-Test and 101113983/Qu-Pilot.

\section*{Author contributions}

J.N., M.K., V.V., O.K., M.B., J.G., J.S. and A.K. conceptualized the work; K.K, K.L and E.T.M. did data curation; K.K., E.M. and K.L. performed formal analysis;  V.V., M.B., J.G., J.S. and A.K. did funding acquisition; K.K., J.N., E.M., P.Sel., T.F., K.L., E.T.M., S.K., S.A., H.S. and A.K. took part in investigation; J.N, P.Sel. and T.F. developed methodology; A.K. did project administration; J.N., P.Sel., T.F., M.R., P.Set., R.L., J.-W.L. and T.R. helped with resources; K.K., E.M., K.L., E.T.M. H.S., V.V., J.G. and A.K. developed software; J.N., V.V., J.G., J.S. and A.K. supervised; K.K., E.M. and K.L visualized the results; K.K., K.L. and A.K. wrote the original draft;  K.K., J.N., E.M, T.F., K.L., M.B. and A.K. did review and editing of the manuscript.

\section*{Data availability}
The data that support the findings of this article are publicly available~\cite{kohopaa_2026_21239237}.


%

\end{document}